\documentclass{ws-ijmpa}
\usepackage[super,compress]{cite}
\usepackage{amssymb}
\usepackage{amsmath}
\usepackage[all]{xy}

\def\lbldef#1#2{\expandafter\gdef\csname #1\endcsname {#2}}
\def\eqn#1#2{\lbldef{#1}{(\ref{#1})}
\begin{equation} #2 \label{#1} \end{equation}}

\def\href#1#2{#2}
                                                                              
\def\tvphi{{q}} 

\begin{document}

\markboth{Jimenez, Talavera \& Verde}
{An effective theory of accelerated expansion}

%
\catchline{}{}{}{}{}
%

\title{An effective theory of accelerated expansion}

\author{Raul Jimenez}
\address{ICREA and
            ICC, Universitat de Barcelona (IEEC-UB), Marti i Franques 1, Barcelona, Spain.\\
            raul.jimenez@icc.ub.edu}

\author{P. Talavera}
\address{ ICC, Universitat de Barcelona (IEEC-UB), Marti i Franques 1, Barcelona, Spain and
          DFEN, Universitat Politecnica de Catalunya, Comte Urgell 187, Barcelona, Spain.}

\author{Licia Verde}
\address{ICREA and ICC, Universitat de Barcelona (IEEC-UB), Marti i Franques 1, Barcelona, Spain \\ liciaverde@icc.ub.edu}

\maketitle

\begin{abstract}
We work out an effective theory of accelerated expansion for the background to describe general phenomena of inflation and acceleration (dark energy) in the Universe. Our aim is to determine from theoretical grounds,  in a physically-motivated and  model independent way, which and how many (free) parameters are needed to broadly capture the physics of a theory describing the background of cosmic acceleration. Our goal is to make as much as possible transparent the physical interpretation of the parameters describing the expansion. We show that, at leading order, there are five independent parameters, of which one can be constrained via general relativity tests. The other four parameters need to be determined by observing and measuring the cosmic expansion rate only, $H(z)$. Therefore we suggest that future cosmology surveys focus on obtaining an accurate as possible measurement of $H(z)$ to constrain the nature of accelerated expansion.
\end{abstract}

\section{Introduction}

It has now been experimentally established \cite{SN,SNII,LSS,Stern,Jimenez} that the
Universe is currently accelerating and present knowledge indicates that eventually will enter a deSitter
phase (see the reviews in Ref.~\cite{Peebles:2002gy,Copeland:2006wr,Sahni:2006pa,Frieman:2008sn} and references therein). There is also experimental evidence that large--scale
structures in the  early Universe were seeded by non-casual perturbations
\cite{wmap01}; the best theoretical model to explain this is inflation,
which consist of an accelerating de-Sitter phase in the  early Universe. Therefore, there is now
strong experimental evidence that the Universe has experienced periods of
accelerated expansion, however we have no satisfactory explanation of
what has been driving it and of the physical mechanism behind it.

Progress can arise both from the theoretical and observational front. It can
happen that a fundamental theory is found that tells us what the nature of
inflation and/or dark energy are. On the other hand, it could be that we
need to exploit astronomical observations in order to zoom in into the detailed properties of the expansion before we can discriminate among competing theories and shed light on the mechanism driving the expansion.

If  the latter scenario is realized the question that arises is: how can we
determine in the most model-independent way, the nature of acceleration? Can this be done in such a way that observational constraints can have an easy and transparent interpretation in terms of the properties of a possible underling physical mechanism? For example, in the context of dark energy, the
most widely used approach is to assume that
the expansion is driven by a scalar field with constant equation of state $p
= w \rho$ and use observations to find $w$ ($p, \rho$ are the pressure and
energy density of the scalar field). Of course, in this framework, a
cosmological constant corresponds to $w=-1$ while scalar fields with
dynamics deviate from this value. It is clear that one cannot assume $w$ to
be constant with time, and that a more general scenario would require a
reconstruction of $w(t)$. It is important to keep in mind that this description is a drastic simplification as it  does not cover all possible scenarios (see e.g.,\cite{Simon}), and  there could be  models where $p, \rho$ are not even defined,  but from observations of any expansion history  one could always reconstruct  an effective $w(t)$. In this case the $w(t)$ would have no physical meaning in terms of properties of a fluid or a scalar field and would be of difficult or ambiguous  theoretical interpretation. Even with this drastic simplification, it is very challenging
observationally to determine $w(t)$ \cite{Simon}. To circumvent this
problem, the community has proposed several parameterizations: some more
closely related to the underlying physics (e.g.,\cite{Simon}) and others
purely phenomenological (\cite{Polarski,Linder} and refs therein). All these parameterizations make
compromises in order to adjust the number of parameters to the observables
in the sky. Even if we do not consider this limitation, it is unclear how
closely related they are to tell us something about the underlying physics
driving the expansion. Moreover degeneracies can be found such that a
constant equation of state can be mimicked by a variable one in these
parameterizations (e.g.,\cite{JimenezLoeb2002}).

Given the above limitations, it is worth asking the question of how to build
a general theory of accelerated expansion that captures as much as possible the physics and thus avoids
 arbitrary parameterizations. This is what we set-up to do in this paper: we build an effective theory for the background of cosmological accelerated expansion. While the philosophy of our approach can be recognized in the works of e.g., Ref~\cite{Weinberg:2008hq,Creminelli,Park:2010cw} we note that our focus is to deal with the background and not with the fluctuations, as was done by the previously referred papers which in turn were inspired by the theory of pions from the 70's. A similar approach has been developed in \cite{Park:2010cw} where the authors have studied thoroughly the phenomenological consequences of their approach. In our approach we propose a physically--motivated way to truncate the effective theory of expansion when there is no physical information about the driver of acceleration, e.g. we do not want to constraint our model to be a quintessence one. In inflation one observes directly the fluctuations but in dark energy we only see the expansion, thus we expect our methodology to be more relevant in the context of dark energy.

\section{A generic set-up}

Our primary goal is to build consistently the scalar potential $V_0(\varphi)$ responsible for the expansion. In order to do so we
shall assume the existence of an energy scale, $\mu$, in the problem where an explicit, but unknown, symmetry is broken \cite{Freese:1990rb}. 
This symmetry breaking will give rise to the potential, which we assume to be almost flat, with small deviations parameterized by the corrections to $V_0(\varphi)$.   This assumption is natural and based on observational grounds: in the case of the current expansion of the Universe this is motivated by the fact that the $\Lambda CDM$ model is a good fit to the data. For inflation, observational constraints indicate that the Hubble parameter is very close to being constant during inflation;  so far there is no evidence for deviations from a flat potential (slow--roll for inflation).
In fact the effective theory approach is built to describe the range of time and scales accessible by observations, not e.g.,  the full potential. Thus, for example, for inflation, it will describe the  $\sim10-20$ e-foldings accessible to observations and for dark energy only the few efoldings since dark energy started affecting the expansion.

The simplest theory of expansion involves, besides gravity, a single scalar field described by the leading Lagrangian density
\eqn{termsp40}{
 { {\cal L}\over  \sqrt{ -g}} = {1\over 2}  A(\varphi) {m_p^2}  R - {1\over 2} B(\varphi)  m^2 g^{\mu\nu} \varphi_{\,,\mu} \varphi_{\,,\nu} -V_0(\varphi)\,,
}
where $m_p$ is the Planck mass and $m$ a parameter with dimensions $[m]={\rm Energy}^2$.  This minimal setup follows from the principle of general covariance and locality for both sectors, gravity and matter.
We have written Eq.~(\ref{termsp40}) as a function of a dimensionless field, $\varphi$, with $A(\varphi), B(\varphi)$
being unknown functions.

\subsection{Power--counting}

The philosophy of our approach, expansion in derivatives (small momenta), can 
be recognized in the works of e.g., Ref.~\cite{Weinberg:2008hq,Creminelli,Park:2010cw}. In addition to these we shall also deal with the explicit symmetry breaking parameters that
set the potential. Eq.~(\ref{termsp40}) is a complete description at leading order and involves generic functions of the dimensionless scalar field $\varphi$. These 
functions could be approximate with truncated expansions once we add in addition a physical motivation. 
In the case of accelerated cosmic expansion it is not however clear a priori how to implement this truncation. 
We thus  begin by motivating our choice of power counting. To do so we shall bear in mind the construction of the low-energy QCD effective theory.
 Although we restrict our argument to the leading-order approximation, this can be extended to higher-order terms.
Contrary to QCD, in a theory of expansion one lacks  a physical observable directly related to the scattering of the scalar field. This has some shortcomings
for the construction of the effective theory:
in low-energy QCD the coefficients modulating the operators in the
Lagrangian density e.g., those equivalent to $A(\varphi)$ and  $B(\varphi)$, can be expanded around its vacuum as an infinite polynomial of pseudo-Goldstone boson fields.
For instance, in the case of QCD, the leading kinetic part will contain the following interaction terms

\begin{eqnarray}
\label{qcdp2}
{\cal L}_{p^2}^{\rm QCD}=\partial_\mu \pi^+ \partial_\mu \pi^-  \hspace*{4cm} && \nonumber \\
+ {1\over 6 F_\pi^2} [-2  \partial_\mu \pi^+ \partial_\mu \pi^- (  \pi^0 \pi^0+ \pi^+\pi^-) \hspace*{1cm} && \nonumber \\
+ \pi^+ \pi^+ \partial_\mu\pi^-\partial_\mu\pi^- + \pi^- \pi^- \partial_\mu\pi^+\partial_\mu\pi^+] +\ldots &&
\end{eqnarray}
where the ellipses stand for terms involving higher powers of the field.
All the terms in (\ref{qcdp2}) have the same weight in the series, ${\cal O}(p^2)$, and therefore they can not be disregarded without any further criteria. In this case the truncation of the polynomial is automatically achieved by computing the contribution to a given scattering $S$ matrix element with a fixed number of external fields \cite{Leutwyler:1993iq}. This pins down a finite number of terms in the expansion. For instance, for the leading two-point function only the first term in (\ref{qcdp2}) contributes while the one between squared brackets contributes to the four-point function. 
Due to the lack of experimental information for the scalar scattering processes in the accelerated cosmic expansion, one can not use the same procedure and without any further assumptions, see for instance \cite{Park:2010cw}, one has to deal with an infinite number of terms at each order of the expansion. A natural way to 
circumvent this situation is to maintain the naive power--counting $[\varphi]={\cal O}(\mu)$. With this (\ref{termsp40}) leads directly to
\eqn{termsp4}{
{ {\cal L}_0\over  \sqrt{ -g}} = {1\over 2} {m_p^2}  R - {1\over 2} g^{\mu\nu} \varphi_{\,,\mu} \varphi_{\,,\nu} -V_0(\varphi)\,.
}
Note that instead Refs. \cite{Weinberg:2008hq,Creminelli} adopted the ``natural" choice $[\varphi]={\cal O}(1)$ equivalent to that in (\ref{qcdp2}). With the latter power--counting the authors of \cite{Park:2010cw} are forced to include extra assumptions in order to gain predictive power.

\subsection{Derivative corrections}
\label{ders}
One of the first corrections to the Lagrangian one can think of are those obtained by increasing the number of derivatives in the scalar sector. 
Those terms are similar to those obtained in a kinetically driven inflation \cite{ArmendarizPicon:1999rj}, and the next-to-leading
contribution can be cast as
\eqn{supp}{
{ \Delta{\cal L} \over  \sqrt{ -g}}  =  f_1  \left( g^{\mu\nu} \varphi_{,\,\mu} \varphi_{,\,\nu}\right)^2\,.
}
The first consequence for the choice of power--counting is that corrections to the scalar kinetic term are a subleading effect starting at $f_1\sim {\cal O}(1/\mu^4)$.
In practice, the change in the power--counting with respect to that presented in  \cite{Weinberg:2008hq}, translates into a rearrangement of the terms in the series (and a different truncation). For instance, the equivalent expression to (\ref{supp}) is \cite{Weinberg:2008hq}
\begin{eqnarray}
\label{suppw}
{\Delta{\cal L} \over   \sqrt{ -g}}  =  f_1(\varphi)  \left( g^{\mu\nu} \varphi_{,\,\mu} \varphi_{,\,\nu}\right)^2 + f_2(\varphi)  g^{\mu\nu} \varphi_{,\mu}
\varphi_{,\nu} \Box\varphi  && \nonumber \\
+f_3(\varphi) \left(\Box\varphi\right)^2 \,. &&
\end{eqnarray}
Comparing with (\ref{supp}) one notices that {\sl i)} we have shifted all the terms to higher orders and {\sl ii)} the terms are not equally weighted and hence 
they contribute to different orders of the expansion. 
As we shall see, sec. (\ref{scapot}), the effects of (\ref{supp}) are subleading to the first corrections to the potential. 

\subsection{Gravity corrections}

Indeed the Einstein-Hilbert action describes quite remarkably well the experimental data up to the visible horizon. If this
fact has to be maintained, once the scalar field is introduced, then its role is a subleading effect and  the interaction 
with the gravitational field is something occurring only at higher orders. Bearing such expectations in mind  we shall  organize the perturbative 
corrections to Eq.~(\ref{termsp4}) involving gravity terms.

Terms involving scalar fields and higher order gravity invariants were already
worked out some time ago \cite{Elizalde:1994sn}, but not in the context of effective field theory, and in the notation and power--counting
of  \cite{Weinberg:2008hq} read
\begin{eqnarray}
\label{termsp6w}
{ \Delta{\cal L}_1\over  \sqrt{ -g}} =
  f_4(\varphi)  R^{\mu\nu} \varphi_{,\,\mu} \varphi_{,\,\nu} + f_5(\varphi)  R g^{\mu\nu}\varphi_{,\,\mu} \varphi_{,\,\nu}  && \nonumber \\
+f_6(\varphi)  R\Box\varphi+f_7(\varphi)  R^2 + f_8(\varphi)  R_{\mu\nu}R^{\mu\nu}  && \nonumber \\
+f_9(\varphi)  C_{\mu\nu\alpha\beta}C^{\mu\nu\alpha\beta}\,, 
\end{eqnarray}
where we neglected parity violating operators. As in the preceding section, implementing the natural power--counting simply
re--orders equally weighted terms to different orders in the expansion and sets 
the coefficients in Eq.~(\ref{termsp6w}) to just numerical factors and not functions of
$\varphi$.  As we have suppressed, via the power--counting, the corrections
due to the scalar field, it is not surprising to obtain that the main effects are due to gravity without the scalar mixing 
\eqn{termsp6}{
{ \Delta{\cal L}_1\over  \sqrt{ -g}} =  
f_7  R^2 + f_8  R_{\mu\nu}R^{\mu\nu} 
+f_9  C_{\mu\nu\alpha\beta}C^{\mu\nu\alpha\beta}\,.
}
The natural order of the constants are $f_7\sim f_8\sim f_9\sim {\cal O}(1)\,,$ but when one computes next-to-leading corrections they acquire a logarithmic dependence on the scale $\mu$. 

At this point we have two {\sl equivalent} ways to handle the problem, which in short amount to 1) drop $f_9$ and effectively include it in $f_7$ and $f_8$ or 2) 
 drop $f_7$ and $f_8$ and keep $f_9$:
\begin{enumerate} 
\item If we were after the equation of motion themselves and not  the perturbations to them,  we could substitute the last term in Eq.~(\ref{termsp6}) involving the Weyl tensor\,,
\begin{eqnarray}
\label{weyl}
C_{\mu\nu\rho\sigma}= R_{\mu\nu\rho\sigma}-{1\over 2} ( g_{\mu\rho} R_{\nu\sigma} - g_{\mu\sigma} R_{\nu\rho} -g_{\nu\rho} R_{\mu\sigma} && \nonumber \\
+g_{\nu\sigma} R_{\mu\rho} ) + {R\over 6} \left( g_{\mu\rho} g_{\nu\sigma} - g_{\nu\rho} g_{\mu\sigma}\right)\,,  &&
\end{eqnarray}
 by a linear combination of curvature bilinears that appears in the Gauss-Bonnet
identity. This, in turn, is a total derivative and, provided our coefficients are simply constants, it does not affect the field equation of motion \cite{Stelle:1976gc}. 
The final result amounts to substitute 
the term modulated by $f_9$  for those of $f_7$ and $f_8$. We would obtain, in this case,
a generalized gravity theory which is equivalent to a multi-scalar-tensor gravity action with four derivatives. It is known, in some specific examples, that the associated vacuum is unstable \cite{Chiba:2005nz}. 
\item As the Weyl tensor vanishes for a de Sitter space-time, and our aim is to describe just perturbations around a flat potential, we find more suitable to work with the $f_9$ term rather than with $f_7$ and $f_8$. As by product this choice allows  us to perform a conformal transformation disentangling the gravity and the scalar field in a neat fashion. 
\end{enumerate}

\subsection{The scalar potential}
\label{scapot}

Up to now the symmetry breaking is not explicitly realized in the Lagrangian construction as its effect only comes in for the construction of the scalar potential term. 
As an essential ingredient we shall demand that the breaking is like in the Higgs mechanism, $\varphi\to\varphi_0+\delta\varphi$, i.e. by shifting the ground state. 
Notice that the derivative terms are unaffected by this shift symmetry.  

It is instructive, once more, to bear in mind the QCD analog case. There the non-derivative part of the Lagrangian has a polynomial form with two essential characteristics: {\sl i)} each term is proportional to 
the explicit symmetry breaking parameter, the pion mass.  {\sl ii)} The field content of each term has chiral symmetry. With this at hand one obtains for the first leading terms 
\begin{eqnarray}
\label{massqcd}
{\cal L}_{{\rm mass}}^{\rm QCD}= -m_\pi^2 \pi^-\pi^+-{1\over 2} m_\pi^2 \pi^0 \pi^0 + {1\over 6 F^2_\pi} m_\pi^2 \pi^+\pi^+ \pi^- \pi^- && \nonumber \\
+ {1\over 6 F^2_\pi} m_\pi^2 \pi^+\pi^- \pi^0 \pi^0+ {1\over 24 F_\pi^2} m_\pi^2 \pi^0\pi^0 \pi^0 \pi^0 && \nonumber \\
+\ldots \hspace*{5.65cm} &&
\end{eqnarray}
with ellipsis indicating terms with increasing numbers of fields {\sl but with the same chiral order.} Notice also that all those terms are local. 
Before proceeding let us stop and explain from where we expect to obtain these local terms in a theory of expansion. If we assume that scalar scattering is mediated,
among others, by gravitons we can integrate out their high modes at this scale, $\mu$. At leading order this is easily implemented in the path integral formalism and is sketched in the figure below 
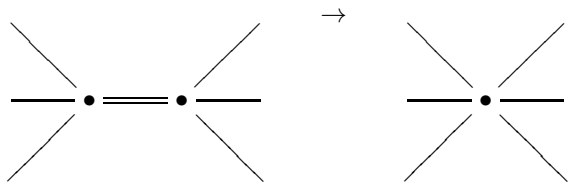
\begin{displaymath}
\xymatrix{
\ar@{-}[dr] & &     & \ar@{-}[dl] \\
\ar@{-}[r] &\bullet &\ar@{=}[l]   \bullet& \ar@{-}[l]\\
\ar@{-}[ur] & &     & \ar@{-}[ul]
}\hspace{0.5cm}\to \hspace{0.5cm}
\xymatrix{
\ar@{-}[dr] & &     \ar@{-}[dl] \\
\ar@{-}[r] &\bullet & \ar@{-}[l]\\
\ar@{-}[ur] & &      \ar@{-}[ul]
}
\end{displaymath}

It is evident that in a theory of expansion we lack all the above information relative to the scales and symmetry breaking pattern.
In the absence of any hint on the symmetries realized by the scalar field we are forced to write the most general polynomial functional for the potential. As in the above case each term will be modulated by a constant, that if different form zero, will signal that a symmetry is broken.
Hence, with full generality we can write
\eqn{vp4}{
V(\varphi) =  \lambda_0+ \lambda_1\varphi + \lambda_2\varphi^2 +\lambda_3\varphi^3 +\lambda_4\varphi^4 + \ldots
}
where the ellipses stand for subleading terms in $1/\mu$. The natural values for the coupling constants in dimension four are
$\lambda_n\sim {\cal O}(\mu^{4-n})$. Therefore any further terms in the series will appear in powers of $1/\mu$ and therefore be supressed. We have included a constant term in the potential that serves as a cosmological constant. 
If some of the rest of the parameters are non-zero this will signal deviations from the $\Lambda CDM$ model.
The stability of the model is ensured at $\varphi\approx 0$ by the analysis of \cite{Bertolami:1987wj}.
Note that choosing some ad-hoc specific values for the parameters the potential can  become that of the chaotic inflation \cite{Linde:1986fc},  the Standard Model Higgs \cite{CervantesCota:1995tz} or the minimal inflation scenario \cite{inf1,inf2}.

Following this logic, we look for terms with mixed products of derivatives with monomials in the scalar field 
\eqn{termsm6}{
{ \Delta{\cal L}_{2} \over  \sqrt{ -g}}  = R (\alpha_1 \varphi+\alpha_2 \varphi^2) + \ldots
}
where $\alpha_1\sim {\cal O}(\mu)\,,\alpha_2\sim {\cal O}(1)$.

\section{The minimal set up}

Collecting the leading  terms from  Eqs.~(\ref{termsp4}), (\ref{termsp6}) and (\ref{termsm6}),  we obtain 
\eqn{born}{
 { {\cal L}\over  \sqrt{- g}} = W(\varphi){m_p^2\over 2}  R +  f_9  C_{\mu\nu\alpha\beta}C^{\mu\nu\alpha\beta} -  {1\over 2}  g^{\mu\nu}  \varphi_{,\,\mu} \varphi_{,\,\nu} -V(\varphi)\,,
}
with 
\eqn{ww}{
W(\varphi) := 1+ {2  \over m_p^2} \alpha_1 \varphi+ {2 \over m_p^2} \alpha_2 \varphi^2\,,
}
 and $V(\varphi)$ given in Eq.(\ref{vp4}).  We expect that if the explicitly symmetry breaking terms are small, the radiative corrections will stay under control keeping the potential sufficiently flat. The addition of these terms has brought the initial Lagrangian density written in the Einstein frame to the final form, Eq.~(\ref{born}), that corresponds to a Lagrangian density in the Jordan frame. This physical system can be mapped onto a  scalar tensor gravity in addition to the field $\varphi$. We do not attribute any special role to the frame in which the theory is formulated \cite{Flanagan:2004bz} 
and for simplicity we  would like to undo this non-minimal coupling to gravity \cite{Faraoni:1998qx}. To do this one performs a conformal transformation at the action level
\cite{Maeda:1988ab}\,, see Appendix,  obtaining as final result the familiar form
\begin{eqnarray}
\label{born4}
 { {\cal S}} = \int d^4\,x \sqrt{- \tilde{g}} \hspace*{5cm} && \nonumber \\
\times  \left( {m_p^2\over 2}  \tilde{R}  
+ f_9  \widetilde{C}_{\mu\nu\alpha\beta} \widetilde{C}^{\mu\nu\alpha\beta}-  {1\over 2}\tilde{g}^{\mu\nu}  \widetilde{\partial}_\mu \tvphi  \widetilde{\partial}_\nu \tvphi - U(\tvphi)\right)\,,
\end{eqnarray}
where $U(\tvphi)$ is given in Eq.~(\ref{newpot}).
In the limit of exact  shift symmetry,  in the  previous expression  the  quantities with tilde become the quantities  before conformal transformation and the potential goes to zero; we thus recover the expression of \cite{Weinberg:2008hq}.
As we are after the first corrections due to the $\lambda_i,\alpha_i, f_i$ coefficients, we discard, after integration, any non-linearities due to them in the potential, obtaining that, from the initial seven free parameters in Eq.~(\ref{born}) ($\lambda_{1,...,4}$, $\alpha_{1,2}$, $f_9$), only five of them must be taken into account: four in the potential, $\lambda_i$,  and one in the pure gravity side $f_9$. All other parameters are subleading or only appear as non-linear corrections. Note that our formalism is an accurate description of any theory of acceleration with one degree of freedom and within a four dimension space-time. Any theory for expansion in more dimensions or that deviates widely from a cosmological constant cannot be accomodated in our current description.

The numerical values for the coefficients $\lambda_i$ and $f_i$ at a given scale are arbitrary and not constrained by symmetry. Their actual value can only be found by matching the theoretical expression for some specific observables with its experimental value. This will be our next task.

\section{Reconstruction procedure}

If we had chosen $f_7$ and $f_8$ over  the $f_9$ term,  in principle constraints could be  obtained  by some pure test of general relativity. For instance,
the E\"ot-Wash experiments provide the strictest bound, $\vert f_7\vert \lesssim 10^{-9} m^2$, while astronomical bounds are weaker  $\vert f_7\vert \lesssim 0.6 \times 10^{18} m^2$ derived from the orbit of Mercury \cite{Berry:2011pb} and $\vert f_7\vert \lesssim 1.7 \times 10^{17} m^2$ from binary pulsars moving in a circular orbit \cite{Naf:2011za}. 
 In cosmological tests of gravity  it is customary to test for deviations from GR at different scales and not to assume any relation between solar system constraints and cosmological constraints.  One of the consequences of our approach is that these parameters  describing modifications of GR, have only a logarithmic dependence on scale. In particular $f_7\sim f_{7,0}+\log(k/\mu)$ where $f_{7,0}$ denotes local measurement and $k\sim 1/r$  with $r$ denoting separation. 
To the best of our knowledge there are no estimates of the values of $f_8$.

Here, we have chosen instead  to consider $f_9$, which could also, in principle, be constrained by tests of gravity. In particular  the $f_9$ term affects the non-linear bispectrum of density perturbations and thus could be constrained from analysis of  forthcoming large-scale structure surveys \cite{Gil-Marin}.

What we have achieved  is a direct  connection between  the parameters describing the physics of the effective theory  and the observables. In fact, let us start from the Friedman equation and the Klein- Gordon equation. To derive them
we assume that the components of $\tilde{g}_{\mu\nu}$ correspond
to that of the physical space that, as usual, would be described by an isotropic, flat homogeneous space-time 
\eqn{frw}{
ds^2 = - N(t)^2 dt^2 + a(t)^2 \delta_{ij} dx^i dx^j\,, 
}
where $N(t)$ is the lapse function and for the moment we set it to unity. We also 
restrict the analysis to classical field configurations that do not break neither homogeneity nor isotropy, since we are interested in describing the background and not the growth of perturbations, $\tvphi=\tvphi(t)$. 
From Eq.~(\ref{born4}) one can obtain the energy-momentum tensor 
\eqn{emt}{
\tilde{G}_{\mu\nu}=\tilde{R}_{\mu\nu}-{1\over 2} \tilde{R}\, \tilde{g}_{\mu\nu} =  \tvphi_{,\,\mu} \tvphi_{,\,\nu}  - \left[ {1\over 2} \tvphi_{,\,\alpha} \tvphi_{,}^{\,\alpha}   
+U(\tvphi) \right]\tilde{g}_{\mu\nu} +T_{\mu\nu}^B,
}
where $T_{\mu\nu}^B$ is the background energy-momentum tensor truncated consistently with the effective order we work. Comparing the matter content of Eq.~(\ref{emt}) with the energy-momentum tensor of a perfect fluid in thermodynamic equilibria one realizes that the pressure and the energy density can be cast as 
$\begin{scriptsize}
 {1\over 2} \tvphi_{,\,\alpha} \tvphi_{,}^{\,\alpha} - U(\tvphi)  = p_\tvphi\,,  {1\over 2} \tvphi_{,\,\alpha} \tvphi_{,}^{\,\alpha} + U(\tvphi)  = \rho_\tvphi\,,
\end{scriptsize}$
respectively.
Notice that the extrema of $p_\tvphi$ with respect to the kinetic energy correspond to the same equation of state as that of a cosmological constant:
$\rho_\tvphi+p_\tvphi=0$.

The Friedman's equations are obtained straightforward once we identify the right hand side of Eq.~(\ref{emt}) as  $ T_{\mu\nu}^q + T_{\mu\nu}^B $. 
The resulting field equations are the standard ones
\eqn{friedman1}{
H^2 = \left({\dot{a}\over a}\right)^2= {1\over 3} \left( \rho_m + \rho_q\right)\,,\quad
{ \ddot{a}\over a}= {1\over 6} (\rho_m+3 p_m+\rho_q+3p_q)\,,
}
and
\eqn{eomsca}{
\ddot{\tvphi} +3 H \dot{\tvphi}-U^\prime=0\,.
}
Dotted quantities stand for derivatives wrt time and prime quantities denote derivatives wrt  the field $q$; $\rho_m$ denotes the matter density parameter and we assume that the Universe is dominated by  collisionless, pressureless  matter,  $p_m=0$. 

In the case of inflation, since $\rho_m=0$ the set of equations simplifies and one reproduces the standard approach. In the case of dark energy  the presence of the matter   density complicates finding a solution. 

Nevertheless, given a set of parameters $\lambda_i$, note that the first Friedman equation and the Klein Gordon  equation can be rewritten in terms of redshift using the fact that $H(z)=-1/(1+z)dz/dt$  and then can be combined in a single differential equation for the field as a function of redshift $q(z)$:
\begin{eqnarray}
\frac{1}{6}\left(\frac{dq}{dz}\right)^2(1+z)^2U' \hspace*{3cm} && \nonumber \\
-\left[U(1+z)+\rho_{m,0}(1+z)^4\right]\frac{dq}{dz}-U'=0, &&
\label{eq:master}
\end{eqnarray}
where $\rho_{m,0}$ denostes the present day matter density.

Once the solution of this equation, $q(z)$ (where only the positive solution is the relevant one) has been obtained, it can be substituted in the expressions for $U'$ and thus of $H(z)$. $H(z)$, being the Universe expansion rate, is the key observable which can be obtained from galaxy surveys via e.g., the standard clocks approach \cite{JimenezLoeb2002,SternHz} or baryon acoustic oscillations measurements (\cite{SeoEisenstein:2003} and references therein).  

In practice however the $\lambda_i$ are not known and  one would like to constrain them from  $H(z)$ measurements, in other words, in practice,  one wants to solve the ``inverse problem" (from $H(z)$ constraints to constraints on $\lambda_i$) while we have shown so far that the ``direct problem" has a solution (from the $\lambda_i$ to the observable $H(z)$).
There are two approaches to solve the ``inverse problem":  {\it i)} Markov Chain Monte Carlo. This is the standard workhorse in cosmology today  for parameter inference.  This problem is perfectly suited to be addressed with this technique. The actual constraints have been presented elsewhere~\cite{Jimenez:2012jg}  {\it ii)}  An exact  analytical approach  which we present below.

\subsection{Analytical solution}

In this approach, and similarly to \cite{Simon}, we use  equivalent quantities to the   inflationary--flow parameters $\{\epsilon_n\}$ \cite{ArmendarizPicon:1999rj} that, in turn, can be cast as functions of $H$ and its derivatives. This approach is both valid for inflation and recent expansion (dark energy). They are defined recursively as
$\begin{scriptsize}
\epsilon_{n+1}={d\log\vert \epsilon_n\vert\over dN}\,,\quad n\ge0\,,
\end{scriptsize}$
$N=\log[a(t)/a(t_0)]$ is the number of efoldings since the time $t_0$ and \ $\epsilon_0:=H(N_0)/H(N)$.

In the case of inflation, although they are theoretical quantities, they can be expressed in terms 
of directly measurable  quantities as the scalar density perturbation index, $n$, and the tensor gravitational perturbation index, $n_g$,
through the relations  $n=1-4 \epsilon_1 - \epsilon_2\,, n_g = -2 \epsilon_2$ \cite{Komatsu:2008hk}. 
In the case of dark energy they can be related to the  Universe expansion (Hubble) rate \cite{Simon}.
For the time being
we shall make use only of the first two parameters $\epsilon_1\,, \epsilon_2$. One can, obviously consider higher-order terms,  although in reality experimental data will be increasingly insensitive to higher derivatives of $H$; thus we refrain from using them and look for other quantities that contain only $\epsilon_1$ and $\epsilon_2$.

We shall split the theoretical input into two classes: the first one which are obtained directly from the definitions of the inflationary--flow parameters
in terms of the potential 
\begin{equation}
\label{eppps}
\begin{array}{c}
\epsilon_1= {m_p^2\over 2} \left( U^\prime\over U\right)^2\,,\quad 
\epsilon_2= {m_p^2} \left( (U^\prime)^2-2 U U^{\prime\prime} \over U^2\right)\,,\quad
\end{array}
\end{equation}
and a second more elaborated class of relations obtained following the lines of \cite{Simon} as we shall outline here.

Using both Friedman equations one can rewrite the first inflationary--flow parameter in terms of the energy density and pressure as
$
\epsilon_1= 1 - {\ddot{a}\over a} H^{-2}= {3\over 2}{\rho_m+\rho_\tvphi+p_m+p_\tvphi\over \rho_m+\rho_\tvphi}\,.
$
The above expression together with the first Friedman equation leads to

\begin{eqnarray}
\label{kf}
{1\over 2} \dot{\tvphi}^2 & = & \epsilon_1 {H^2\over \kappa^2}-{1\over 2}(\rho_m+p_m), \\ \nonumber
U(q) & = & (3-\epsilon_1) {H^2\over \kappa^2} +{1\over 2}(\rho_m-p_m), \\ \nonumber
\kappa^2  & = & {8\pi\over m_p^2}. 
\end{eqnarray}

Integrating the former of these relations with respect to time we can obtain the field in terms of the observables. 

The last relation we need makes use of the scalar equation of motion. Rewriting Eq.~(\ref{eomsca}) as $U^\prime = -(3H\dot{q}^2+\dot{q}\ddot{q})/\dot{q}$
and using Eq.~(\ref{kf}), the derivative of the potential wrt the field becomes
\begin{eqnarray}
\label{Uder}
U^\prime(q) = - {3\over 2\sqrt{\pi}} m_p H^2 \epsilon_1^{1/2} \left[ 1 -{\kappa^2\over 2 H^2\epsilon_1}(\rho_T+p_T)\right]^{-1/2} & & \nonumber\\
\times \left\{
1-{\epsilon_1\over 3}  +  {\epsilon_2\over 6}   -{\kappa^2\over 6 H^3 \epsilon_1 }\left[ 3 H (\rho_T + p_T)  + {1\over 2} (\dot{\rho}_T+\dot{p}_T) \right] 
\right\}. & &
\end{eqnarray}
where $\rho_T$ and $p_T$ describe the energy density and pressure contribution from higher derivative curvature terms. The above expression together with Eq.~(\ref{kf}) and (\ref{eppps}) determine the system of five equations from where we determine the set \\
$\{\lambda_1\,,\ldots\,,\lambda_4,q(t)\}$. 
Compared with the ``Monte Carlo" approach where only  the $H(z)$ determination is needed, in this approach one needs to determine observationally $H$, $dH/dz$ and $d^2H/dz^2$ which is very challenging. 

Although an analytical treatment for the full system can be done, the results are too cumbersome to be presented in a reasonably manageable way. Instead we report an implicit system of equations (where we have used $m_p = 1$):

\begin{eqnarray}
\label{lambss}
&& \lambda_1 = 4\sqrt{2} {q^2(2\epsilon_1-\epsilon_2)-4\over 16+q^4(2\epsilon_1-\epsilon_2)^2 } 
q^3 \epsilon_1^{1/2} (\lambda_3 + 3 q \lambda_4) \nonumber  \\ 
&& + {16q^2 (3\lambda_3+8q\lambda_4)\over 16+q^4(2\epsilon_1-\epsilon_2)^2} \nonumber\\
&&\lambda_2 =  -4\sqrt{2}{q^4 \epsilon_1^{1/2} (\lambda_3 + 3 q \lambda_4)(2\epsilon_1-\epsilon_2) \over 16+q^4 (2\epsilon_1-\epsilon_2)^2-8q^2(2\epsilon_1+\epsilon_2)}  \nonumber\\
&&+4 q{  \left[ q^2(14\epsilon_1+5\epsilon_2)-12 \right] \lambda_3 + 3 q \left[  q^2(10\epsilon_1 + 3 \epsilon_2)-8 \right] \lambda_4 \over 16 + q^4 (2\epsilon_1-\epsilon_2)^2 - 8 q^2 (2\epsilon_1+\epsilon_2)}  \nonumber\\
&&\lambda_3 = {3\over 8\sqrt{\pi} q^2} {\dot{p}_m+\dot{\rho}_m+6 H (p_m+\rho_m)\over p_m+\rho_m-2H^2\epsilon_1} \epsilon_1^{-1/2} \nonumber \\
&&(4H^2\epsilon_1-2p_m-2\rho_m)^{1/2}
+ {3\over \sqrt{2\pi} q^2} H^3 \epsilon_1^{1/2} \nonumber \\
&&+{1\over q^3}\left\{2 p_m -3 q \lambda_1 -2 [ 2H^2 (\epsilon_1-3)+q^2 \lambda_2 +\rho_m]\right\}  \nonumber\\
&&\lambda_4 =  {1\over 4\sqrt{2\pi} q^3}  \nonumber\\
&&{3\dot{p}_m+3\dot{\rho}_m+2 H [H^2 \epsilon_1(2\epsilon_1 -\epsilon_2-6) + 9p_m+9\rho_m]\over \epsilon_1^{1/2}(2H^2\epsilon_1-p_m-\rho_m)^{1/2}}  \nonumber\\
&&+ {3\over q^2} H^2 \epsilon_1+ {1\over 2 q^4} (4 q \lambda_1 + 2 q^2 \lambda_2 + 3\rho_m-18H^2-3p_m),  \nonumber\\
\end{eqnarray} 
where $\rho_m$ and $p_m$ are the density and pressure of any matter species present during the accelerating phase.

\section{Constraints on the parameter space}

Although it might look that there is total freedom in choosing the values for the unknown parameters this is not the case. There are several consistency relations that the
effective theory must fulfill; these restrict the range of the parameters numerical values.

\subsection{The fate of gravitational higher-order corrections}

One particular case of accelerated expansion is inflation. This is believed to be a semiclassical effect. In that case quantum gravity corrections can be neglected because they are small in the following sense: curvature corrections must be suppressed at the Planck scale. This is not always the case and in fact some inflationary models based on pseudo-Goldstone fields 
suffer from this pathology. We do not dwell more on this issue but instead estimate roughly some constraints that the pure gravitational part must fulfill in order to satisfy our perturbative expansion. In particular, we shall demand that the contribution to the energy density coming from higher orders terms is subleading. For that  purpose we focus on the pure gravitational part of Eq.~(\ref{emt}) viz. $ \tilde{G}_{\mu\nu} -T_{\mu\nu}^B=0$ with 

\begin{eqnarray}
\label{hdergrav}
- T_{\mu\nu}^B =&& 2 f_7 R \left( \tilde{R}_{\mu\nu}-{1\over 4}  \tilde{R}  \tilde{g}_{\mu\nu}\right) + 2 f_7 ( \tilde{g}_{\mu\nu} \Box  \tilde{R} - \nabla_\mu\nabla_\nu  \tilde{R}) \nonumber \\
&& + 2 f_8\left(  \tilde{R}_{\mu\rho}  \tilde{R}^\rho_\nu-{1\over 4}  \tilde{R}^{\rho\sigma} \tilde{R}_{\rho\sigma}  \tilde{g}_{\mu\nu}\right) \nonumber
\\&&\hspace{-0.7cm}+ f_8\left( \Box  \tilde{R}_{\mu\nu}+{1\over 2} \Box  \tilde{R}  \tilde{g}_{\mu\nu}-2\nabla_\mu\nabla_{(\rho}^{~~}
 \tilde{R}_{\nu)}^{~~\rho} \right)\,.
\end{eqnarray}

Under the previous requirement of absolute convergence for the effective series the $\{0,0\}$ component leads to the inequality
\begin{eqnarray}
\label{hen00}
&& (26 f_7 -6f_8) \dot{H}^2 +2(4f_7-f_8) \dddot H + 2 (16f_7-f_8) H  \ddot{H} \nonumber \\
&& -6(2f_7+f_8) H^2 \dot{H}\ll H^2\,.
\end{eqnarray}
If observational constraints can be obtained from cosmological observations that lead to values of $H, \dot H, \ddot H, \dddot H$, then, from the above equation we can constrain the values of $f_7$ and $f_8$.

\subsection{Inflation}
Many inflation models require a  transplankian scale for the scalar field  $q\gg m_p$.
To circumvent this problem some non-minimal scalar-gravity coupling has been advocated recently \cite{Germani:2010gm}. Although this approach is promising for Higg-like potential models, we want to stress that it can not be implemented in our effective field theory because {\sl it mixes gravitational terms of different effective orders}  within our power-counting set up.

In order to ascertain the correctness of the framework developed in this paper, we check whether this situation can be present. 
Using Eq.~(\ref{frw}) into Eq.~(\ref{born4} ) and varying the action with respect to the lapse function and then setting it afterwards to $1$ by time reparameterization invariance, 
we obtain the Hamiltonian constraint $H^2 = {1\over 6 m_p^2} \left( \dot{q}^2 + 2 U\right)$. The slow--roll conditions can be read automatically from the previous 
expression and Eq.~(\ref{eomsca}): $\dot{q}^2 \ll 2 U$ and $\vert \ddot{q}\vert^2 \ll 3 H \vert\dot{q}\vert$. These two expressions imply 
$H^2\approx {1\over 3 m_p^2} U\,,\dot{q} \approx - {V^\prime\over 3 H}$ from where one can deduce
\eqn{slrbreak}{
-{\dot{H}\over H^2} \approx {m_p^2\over 2 q^2} \left({\lambda_1 +2 \lambda_2 q +3 \lambda_3 q^2 + 4 \lambda_4 q^3\over 
\lambda_1 +\lambda_2 q + \lambda_3 q^2 +  \lambda_4 q^3}\right)^2 \ll 1\,.
}
The standard constraint, for example in chaotic inflation, to obtain enough number of efoldings, implies that the inequality can be only achieved if $q \gg m_p$. However, from the above equation, it is now clear how to obtain sub-planckian inflation, for instance, $ - \lambda_1\approx\lambda_2\approx\lambda_3 \gg \lambda_4$ will produce enough number of efoldings and  inflate without having to deal with transplanckian scales, $q\ll m_p\,.$ 

\section{Conclusions}

We have developed an effective theory of the background of accelerated expansion with the aim of  making a direct connection  between  cosmological observables and  parameters  describing  the physics behind the expansion.
We have adopted what we propose as the ``natural" power counting to expand (and truncate)  the expression for the Lagrangian describing the accelerating Universe. Our choice is motivated by being natural for scalar fields.
 
In doing so we have discovered that only five parameters are needed to fully described the (effective) theory at leading order. These parameters are readily connected to the physics behind expansion and thus are the most natural set an observer should  attempt to determine. In particular one parameter  describes deviations from standard --general relativity-- gravity and the other  four completely determine the expansion history. Although our treatment is general to both inflation   and dark energy, we have concentrated on the case for dark energy which yields less standard results.

 Our conclusion is  that one should do observations that test deviations from general relativity (via e.g., constraints on the growth of cosmological structure or higher-order correlations of the matter  density field), and measure the Hubble parameter $H(z)$ with  the best possible  accuracy in at least four redshift bins. Once $H(z)$ has been measured, Eq.~ (\ref{eq:master}) can be used to obtain the physical parameters of the Lagrangian that describes the accelerated expansion.

Note that, even neglecting modifications to General Relativity, if from observations one wanted to  reconstruct in a non-parametric way the potential of the accelerating field, both the Hubble parameter  $H(z)$ and is derivative $dH(z)/dt$ would be needed as shown in \cite{Simon}. Here, the effective theory expansion already provides a parameterization of the potential as a function of the field, and thus only a measurement of $H(z)$ and  the matter density are needed.

A priori we have no restrictions on the actual values of the unknowns in the theory, beside imposing the validity of the effective theory expansion. If, as an outcome of  confronting the theory with  experimental data in Eq.~ (\ref{eq:master}) some of these parameters turns to be numerically suppressed, we would have learned something about  the pattern of the explicit symmetry breaking for the scalar field. If, on the contrary,  it turns out their numerical values do not match the a priori power--counting estimates,  this would be an indication of a break down of the proposed approach and that an alternative power--counting must be implemented. 
Although we have tackled only the leading contribution, perturbations on any of both fields can be computed in a straightforward manner  along  the lines of Ref.~\cite{Weinberg:2008hq}. For instance, the gravitational corrections turns to be identical to those found by Ref.~\cite{Weinberg:2008hq}.
In our case the speed of sound equals unity, $c_s=1$. Corrections to this value come from higher order operators as in Eq.~(\ref{supp}).

Future cosmology surveys promise to provide measurements of the expansion rate --either directly measuring H(z), or closely related quantities such as the angular  diameter distance  or the luminosity distance as a function of redshift-- with percent precision over a wide range of redshift ($z \sim 3$)  and over tens of redshift bins. This will offer a unique opportunity to gain insight   into the mechanism of cosmic acceleration. In recent work \cite{Jimenez:2012jg} we have used the newly determined $H(z)$ data by Ref.~\cite{Moresco:2012jh} to put constraints on the effective potential 
of dark energy and found it to be consistent with a flat potential, with an allowed deviation of 6\% from flat (Fig.~3 in \cite{Jimenez:2012jg}). Further, we showed how the field has moved in most of the models constraints, thus illustrating the power of the effective field theory approach to elucidate the nature of accelerated expansion: if the fact that the field has moved is confirmed by future and more accurate measurements of $H(z)$, this would indicate that dark energy was caused by a  pseudo-goldstone boson and therefore it was the result of a symmetry breaking. In this paper we have concentrated on the background expansion and neglected perturbations. While this is satisfactory for dark energy, as the observational perspectives of detecting dark energy fluctutations are very gloomy, it is not for inflation, where perturbations are relevant.

\section*{Appendix}

In this appendix we give the explicit steps that leads to Eq.~\ref{born4bis}.
The conformal transformation takes the form
$
g_{\mu\nu}\to\tilde{g}_{\mu\nu}=\Omega^2 g_{\mu\nu} \,$ where $\Omega^2=W(\varphi)$. 
That 
leads to the Lagrangian density
\begin{eqnarray}
\label{born3}
 \!\!\!\!\!\!\!\!\!\!\!  { {\cal L}\over  \sqrt{- \tilde{g}}}  =   {m_p^2\over 2}  \tilde{R}  
+ f_9  \widetilde{C}_{\mu\nu\alpha\beta} \widetilde{C}^{\mu\nu\alpha\beta}  \hspace*{2cm} && \nonumber \\
 -  {1\over 2}\left[{1\over \Omega^2}+{6 (\alpha_1+ 2 \alpha_2 \varphi)^2 \over m_p^2\Omega^4}\right]  \tilde{g}^{\mu\nu}  \widetilde{\partial}_\mu \varphi  \widetilde{\partial}_\nu \varphi - {V(\varphi)\over \Omega^4}\,, &&
\end{eqnarray}
where we have used the fact that under the action of the conformal transformation the Weyl tensor transform as
$C_{\mu\nu\alpha}^{~~~~\beta} =  \widetilde{C}_{\mu\nu\alpha}^{~~~~\beta} $.

One can make a further step and normalize the kinetic term for the scalar field in Eq.~(\ref{born3}) using the differential redefinition  
\eqn{sfidif}{
\Omega^4 \left({d\tvphi\over  d\varphi}\right)^2=\Omega^2+{6\over m_p^2}(\alpha_1 + 2 \alpha_2 \varphi)^2\,.
}
With this one finally obtains the familiar form
\begin{eqnarray}
\label{born4bis}
 { {\cal S}} = \int d^4\,x \sqrt{- \tilde{g}}  \hspace*{5cm} && \nonumber \\
\times  \left( {m_p^2\over 2}  \tilde{R}  
+ f_9  \widetilde{C}_{\mu\nu\alpha\beta} \widetilde{C}^{\mu\nu\alpha\beta}-  {1\over 2}\tilde{g}^{\mu\nu}  \widetilde{\partial}_\mu \tvphi  \widetilde{\partial}_\nu \tvphi - U(\tvphi)\right)\,, &&
\end{eqnarray}
with 
\eqn{newpot}{
U(\tvphi) := {V(\varphi)\over \Omega^4}\,.
}
In the limit of exact  shift symmetry,  in the  previous expression  the  quantities with tilde become the quantities  before conformal transformation and the potential goes to zero; we thus recover the expression of Ref.~\cite{Weinberg:2008hq}.
Thus the $\alpha_i$ parameters can be seen as the responsible for the spin-0 deformations of the metric; if one writes the corrections to the metric field as
$\tilde{g}_{\mu\nu} = \Omega^2 g_{\mu\nu}\simeq \eta_{\mu\nu} + \rho_{\mu\nu}= (1+\Omega^2)\eta_{\mu\nu}+h_{\mu\nu}$, the canonical action for the spin-2 field 
is not obtained from the field $h_{\mu\nu}$, but it is instead given by $\rho_{\mu\nu}$\,, i.e., the conformal transformation mixes the degrees of freedom corresponding to the scalar and the tensor modes.
As mentioned above, we expect the parameters $\alpha_i$ to be small quantities implying that the spin-0 deformations are due to the symmetry breaking.
In addition small values ensure the stability of the solution \cite{Carvalho:2004ty}.

\section*{Acknowledgements}

We would like to thank  Luis Alvarez-Gaume \& Jorge Nore\~na for discussions.  We want to specially thank V. Faraoni for bringing to our attention
some relevant references. The research of PT\
was supported in part by grant FPA2010-20807,
 grant 2009SGR502  and the
grant Consolider CPAN.
LV is supported by FP7- IDEAS Phys.LSS 240117.
LV and RJ are supported by MICINN grant AYA2008-03531


\begin{thebibliography}{90}

\bibitem{SN}
  S.~Perlmutter {\it et al.},
  Astrophys.\ J.\  {\bf 517} (1999) 565
  [arXiv:astro-ph/9812133].
  
\bibitem{SNII}  
  A.~G.~Riess {\it et al.},
  Astron.\ J.\  {\bf 116} (1998)  1009
  [arXiv:astro-ph/9805201].

\bibitem{LSS} 
  W.~J.~Percival {\it et al.}  [The 2dFGRS Collaboration],
  Mon.\ Not.\ Roy.\ Astron.\ Soc.\  {\bf 327} (2001) 1297
  [arXiv:astro-ph/0105252].
  
\bibitem{Stern} Stern, D., Jimenez, R., Verde, L., Kamionkowski, M., Stanford, S.~A.\ 2010. Journal of Cosmology and Astro-Particle Physics 2, 8. 

\bibitem{Jimenez} Jimenez, R., Verde, L., Treu, T., Stern, D.\ 2003.\  The Astrophysical Journal 593, 622.

\bibitem{Peebles:2002gy}
  P.~J.~E.~Peebles, B.~Ratra,
  Rev.\ Mod.\ Phys.\  {\bf 75 } (2003)  559-606.
  [astro-ph/0207347].

\bibitem{Copeland:2006wr}
  E.~J.~Copeland, M.~Sami, S.~Tsujikawa,
  Int.\ J.\ Mod.\ Phys.\  {\bf D15 } (2006)  1753-1936.
  [hep-th/0603057].     

\bibitem{Sahni:2006pa}
  V.~Sahni and A.~Starobinsky,
  Int.\ J.\ Mod.\ Phys.\ D {\bf 15} (2006) 2105
  [astro-ph/0610026].

\bibitem{Frieman:2008sn}
  J.~Frieman, M.~Turner and D.~Huterer,
  Ann.\ Rev.\ Astron.\ Astrophys.\  {\bf 46} (2008) 385
  [arXiv:0803.0982 [astro-ph]].
    
\bibitem{wmap01} Peiris, H.~V., et al. 2003.\ The Astrophysical Journal Supplement Series 148, 213. 

\bibitem{Simon}
  J.~Simon, L.~Verde, R.~Jimenez,
  Phys.\ Rev.\  {\bf D71 } (2005)  123001.
  [astro-ph/0412269].

\bibitem{Polarski} Chevallier, 
M., Polarski, D.\ 2001.\ International Journal of Modern Physics D 10, 213.

\bibitem{Linder} Linder, E.~V.\ 2003.\ Physical Review Letters 90, 091301.

\bibitem{JimenezLoeb2002}
  R.~Jimenez, A.~Loeb,
  Astrophys.\ J.\  {\bf 573 } (2002)  37-42.
  [astro-ph/0106145].


\bibitem{Weinberg:2008hq}
  S.~Weinberg,
  Phys.\ Rev.\  {\bf D77}, 123541 (2008).
  [arXiv:0804.4291 [hep-th]].

\bibitem{Creminelli}Cheung, C., Fitzpatrick, 
A.~L., Kaplan, J., Senatore, L., 
\& Creminelli, P.\ 2008, Journal of High Energy Physics, 3, 14 

\bibitem{Park:2010cw}
  M.~Park, K.~M.~Zurek and S.~Watson,
  Phys.\ Rev.\ D {\bf 81} (2010) 124008
  [arXiv:1003.1722 [hep-th]].

\bibitem{Freese:1990rb} {\sl For an explicit model see for instance:}
  K.~Freese, J.~A.~Frieman, A.~V.~Olinto,
  Phys.\ Rev.\ Lett.\  {\bf 65 } (1990)  3233-3236.  
 
\bibitem{Leutwyler:1993iq}
 {\it See for instance:}  H.~Leutwyler,
  Annals Phys.\  {\bf 235} (1994) 165
  [arXiv:hep-ph/9311274].

\bibitem{ArmendarizPicon:1999rj}
  C.~Armendariz-Picon, T.~Damour, V.~F.~Mukhanov,
  Phys.\ Lett.\  {\bf B458}, 209-218 (1999).
  [hep-th/9904075].

\bibitem{Komatsu:2008hk}
  E.~Komatsu {\it et al.} [ WMAP Collaboration ],
  Astrophys.\ J.\ Suppl.\  {\bf 180 } (2009)  330-376.
  [arXiv:0803.0547 [astro-ph]].  



\bibitem{Elizalde:1994sn}
  E.~Elizalde, A.~G.~Zheksenaev, S.~D.~Odintsov and I.~L.~Shapiro,
  Phys.\ Lett.\  B {\bf 328} (1994) 297
  [arXiv:hep-th/9402154].


\bibitem{Stelle:1976gc}
  K.~S.~Stelle,
  Phys.\ Rev.\  {\bf D16 } (1977)  953-969.


\bibitem{Chiba:2005nz}
  T.~Chiba,
  JCAP {\bf 0503} (2005) 008
  [arXiv:gr-qc/0502070].

\bibitem{Bertolami:1987wj}
  O.~Bertolami,
  Phys.\ Lett.\  {\bf B186 } (1987)  161-166.
  

\bibitem{Linde:1986fc}
  A.~D.~Linde,
  Mod.\ Phys.\ Lett.\  {\bf A1 } (1986)  81.


\bibitem{CervantesCota:1995tz}
  J.~L.~Cervantes-Cota, H.~Dehnen,
  Nucl.\ Phys.\  {\bf B442 } (1995)  391-412.
  [astro-ph/9505069].

\bibitem{inf1} 
{\'A}lvarez-Gaum{\'e}, L., G{\'o}mez, C., Jimenez, R.\ 2010.\ Physics Letters B 690, 68-72. 

\bibitem{inf2} 
{\'A}lvarez-Gaum{\'e}, L., G{\'o}mez, C., Jimenez, R.\ 2011.\ Journal of Cosmology and Astro-Particle Physics 3, 27. 

\bibitem{Flanagan:2004bz}
  E.~E.~Flanagan,
  Class.\ Quant.\ Grav.\  {\bf 21 } (2004)  3817.
  [gr-qc/0403063].  


\bibitem{Faraoni:1998qx}
  V.~Faraoni, E.~Gunzig, P.~Nardone,
  Fund.\ Cosmic Phys.\  {\bf 20 } (1999)  121.
  [gr-qc/9811047].


\bibitem{Maeda:1988ab}
  K.~i.~Maeda,
  Phys.\ Rev.\  D {\bf 39}, 3159 (1989).

\bibitem{Berry:2011pb}
  C.~P.~L.~Berry, J.~R.~Gair,
  Phys.\ Rev.\  {\bf D83 } (2011)  104022.
  [arXiv:1104.0819 [gr-qc]].
  
\bibitem{Naf:2011za}
  J.~Naf, P.~Jetzer,
  [arXiv:1104.2200 [gr-qc]].


\bibitem{Gil-Marin}
  H.~Gil-Marin, F.~Schmidt, W.~Hu, R.~Jimenez and L.~Verde,
  JCAP {\bf 1111} (2011) 019
  [arXiv:1109.2115 [astro-ph.CO]].

\bibitem{SternHz} Stern, D., Jimenez, R., Verde, L., Kamionkowski, M., \& Stanford, S.~A.\ 2010, JCAP, 2, 8 

\bibitem{SeoEisenstein:2003} Seo, H.-J., \& Eisenstein, D.~J.\ 2003, ApJ, 598, 720 

%
%


\bibitem{Germani:2010gm}
  C.~Germani, A.~Kehagias,
  Phys.\ Rev.\ Lett.\  {\bf 105 } (2010)  011302.
  [arXiv:1003.2635 [hep-ph]].  
  
\bibitem{Jimenez:2012jg}
  R.~Jimenez, P.~Talavera, L.~Verde, M.~Moresco, A.~Cimatti and L.~Pozzetti, JCAP (2012), 3, 14,
  arXiv:1201.3608 [astro-ph.CO].
  
\bibitem{Moresco:2012jh}
  M.~Moresco, A.~Cimatti, R.~Jimenez, L.~Pozzetti, G.~Zamorani, M.~Bolzonella, J.~Dunlop and F.~Lamareille {\it et al.}, JCAP (2012), 8, 6,
  arXiv:1201.3609 [astro-ph.CO].

\bibitem{Carvalho:2004ty}
  F.~C.~Carvalho, A.~Saa,
  Phys.\ Rev.\  {\bf D70 } (2004)  087302.
  [astro-ph/0408013].
  
  
       
\end{thebibliography}
\end{document}